\documentclass{emulateapj}
\usepackage{amssymb, amsmath, epsfig}

\newcommand{\Msun}{{M_{\odot}}}

\newcommand{\HI}{H{\sc ~i}}
\newcommand{\HII}{H{\sc ~ii}}
\newcommand{\HeI}{He{\sc ~i}}
\newcommand{\HeII}{He{\sc ~ii}}

\begin{document}

\title{On Lyman-limit Systems and the Evolution of the Intergalactic Ionizing Background}
\author{Matthew McQuinn\altaffilmark{1}, S. Peng Oh\altaffilmark{2} , Claude-Andr{\'e} Faucher-Gigu{\`e}re\altaffilmark{1}}

\altaffiltext{1} {Department of Astronomy, University of California, Berkeley, CA 94720, USA; mmcquinn@berkeley.edu\\}
\altaffiltext{2} {Department of Physics, University of California, Santa Barbara, CA 93106, USA}

\begin{abstract} 
We study the properties of self-shielding intergalactic absorption systems and their implications for the ionizing background.
We find that cosmological simulations post-processed with detailed radiative transfer calculations generally are able to reproduce the observed abundance of Lyman-limit systems, and we highlight possible discrepancies between the observations and simulations.  This comparison tests cosmological simulations at overdensities of $\sim100$.  Furthermore, we show that the properties of Lyman-limit systems in these simulations, in simple semi-analytic arguments, and as suggested by recent observations indicate that a small change in the ionizing emissivity of the sources would have resulted in a much larger change in the amplitude of the intergalactic \HI-ionizing background (with this scaling strengthening with increasing redshift).  This strong scaling could explain the rapid evolution in the Ly$\alpha$ forest transmission observed at $z\approx6$.  Our calculations agree with the suggestion of simpler models that the comoving ionizing emissivity was constant or even increasing from $z=3$ to $6$.  Our calculations also provide a more rigorous estimate than in previous studies for the clumping factor of intergalactic gas after reionization, which we estimate was $\approx 2-3$ at $z=6$.
\end{abstract}


\section{introduction}

The hydrogen between galaxies was reionized more than $12.5~$Gyr ago \citep{fan06}.  After this event, a largely uniform 
\HI-ionizing radiation background pervaded the intergalactic medium (IGM) and kept the
hydrogen highly ionized everywhere except within rare, 
dense pockets \citep{gunn65, cen94, miralda96, hernquist96, haardt96, katz96}.  The amplitude of this background appears to have declined quickly above a redshift of $z=6$ (e.g., \citealt{fan06}) and to have stayed relatively constant over $2<z<4$ (e.g., \citealt{faucher08b}). It is of some debate whether this decline owed to the overlap stage of cosmological reionization or to something more mundane (e.g., \citealt{gnedin00, fan06, becker07}).  In addition, the dense self-shielding systems that remained in the IGM, called Lyman-limit systems, provide a window into structure formation in a different density regime than explored by other large-scale structure probes.

This paper aims to study intergalactic radiative transfer and the properties of Lyman-limit systems (systems  with \HI\ columns of $10^{17.2} < N_{\rm HI} < 10^{19}~$cm$^{-2}$).  To do so, we post-process cosmological simulations (both with and without galactic feedback implementations) with ionizing radiative transfer.  The abundance of Lyman-limit systems in the post-processed simulations is then compared with measurements of their abundance from quasar absorption line studies.  This comparison tests how cosmological simulations fare at the outskirts of galactic halos.  Previous attempts to model Lyman-limit systems in simulations had reported various levels of agreement with observations and had used much smaller box sizes and particle numbers than are presently feasible \citep{katz96, gardner01, kohler07}.  In addition, recent observations have significantly improved the constraints on the abundance of Lyman-limit systems \citep{prochaska09, prochaska10, songaila10}. 

Our calculations also provide an estimate for the \HI-ionizing emissivity of star-forming galaxies and quasars since, after cosmological reionization, the production rate of ionizing photons was in balance with the number of absorptions.  These absorptions occurred primarily within Lyman-limit systems.  Our determinations of the emissivity using the simulations agree with observations at redshifts where it can be measured observationally ($2 < z<4$), and they also provide a means to estimate the emissivity at higher redshifts.  We also study the relationship between the emissivity of ionizing photons and the intensity of the ionizing background.  Interestingly, we find that small changes in the emissivity at high redshifts could have resulted in much larger changes in the ionizing background.  This finding has interesting implications for the observed trends in this background.


 

This paper is organized as follows.  Section~\ref{sec:thecase} describes the observations and our numerical techniques.  Section~\ref{sec:comparison} compares our numerical calculations of $\partial^2{\cal N}/\partial z\,\partial N_{\rm HI}$  -- the number of absorption systems per unit redshift per \HI\ column ($N_{\rm HI}$) -- with observations.  Section~\ref{sec:relationship} studies the relationship between dense absorption systems, the ionizing emissivity, and the ionizing background.  Section~\ref{sec:other} considers how physical effects that may not be captured in the simulations could alter our conclusions.  Finally, Appendix~\ref{ap:properties} provides supplementary information regarding the properties of the simulated Lyman-limit systems.  Our calculations assume a flat $\Lambda$CDM cosmology with $(\Omega_m,\Omega_b,h,\sigma_8,n_s) = (0.28,0.046,0.70,0.82,0.96)$ and a hydrogen mass fraction of $0.75$, consistent with the latest Wilkinson Microwave Anisotropy Probe analysis \citep{komatsu10}.\\

\section{Observations and Models of \HI\ Absorbers}
\label{sec:thecase}

\begin{figure*}
\begin{center}
{\epsfig{file=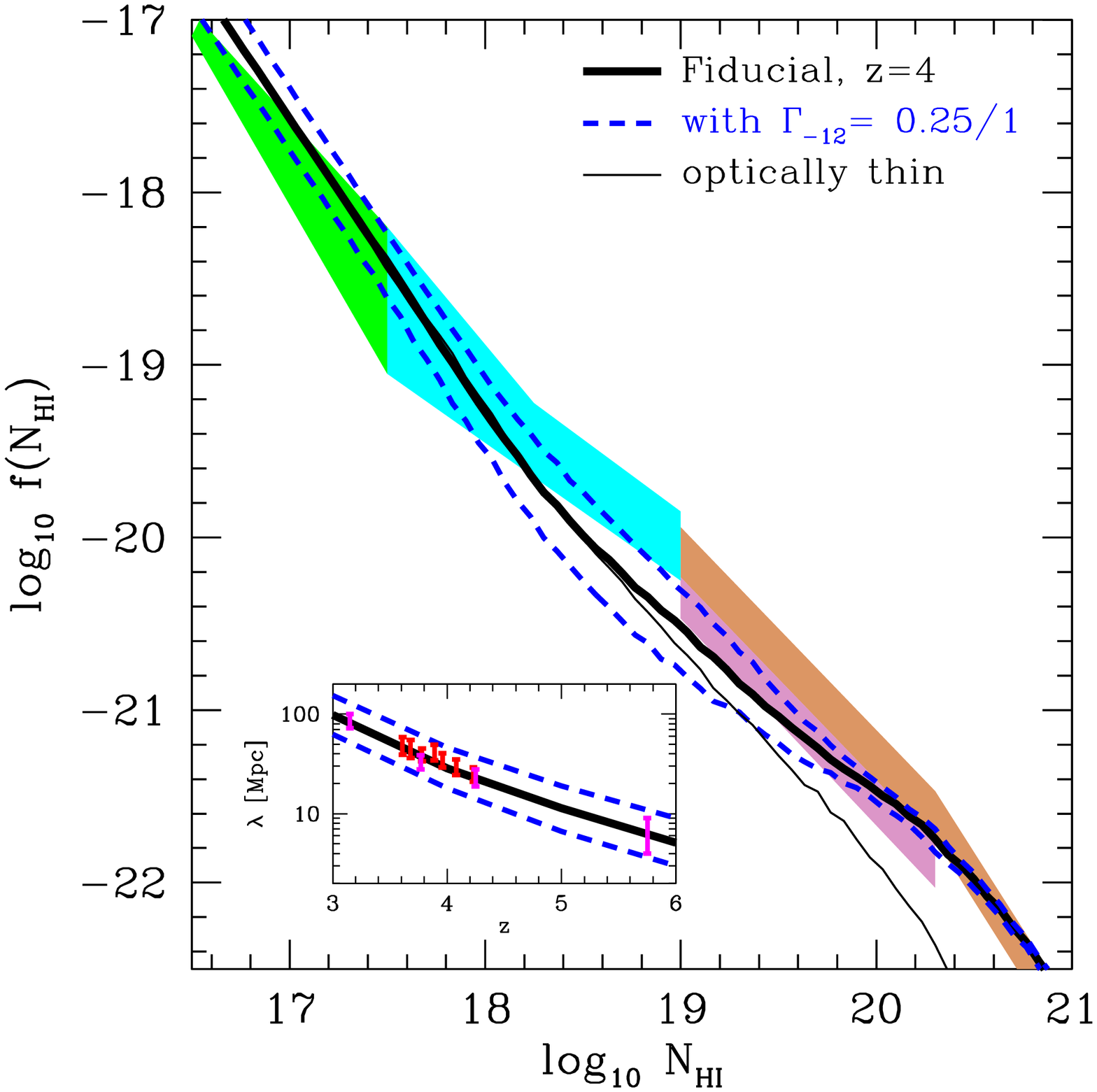, width=8.5cm}}{\epsfig{file=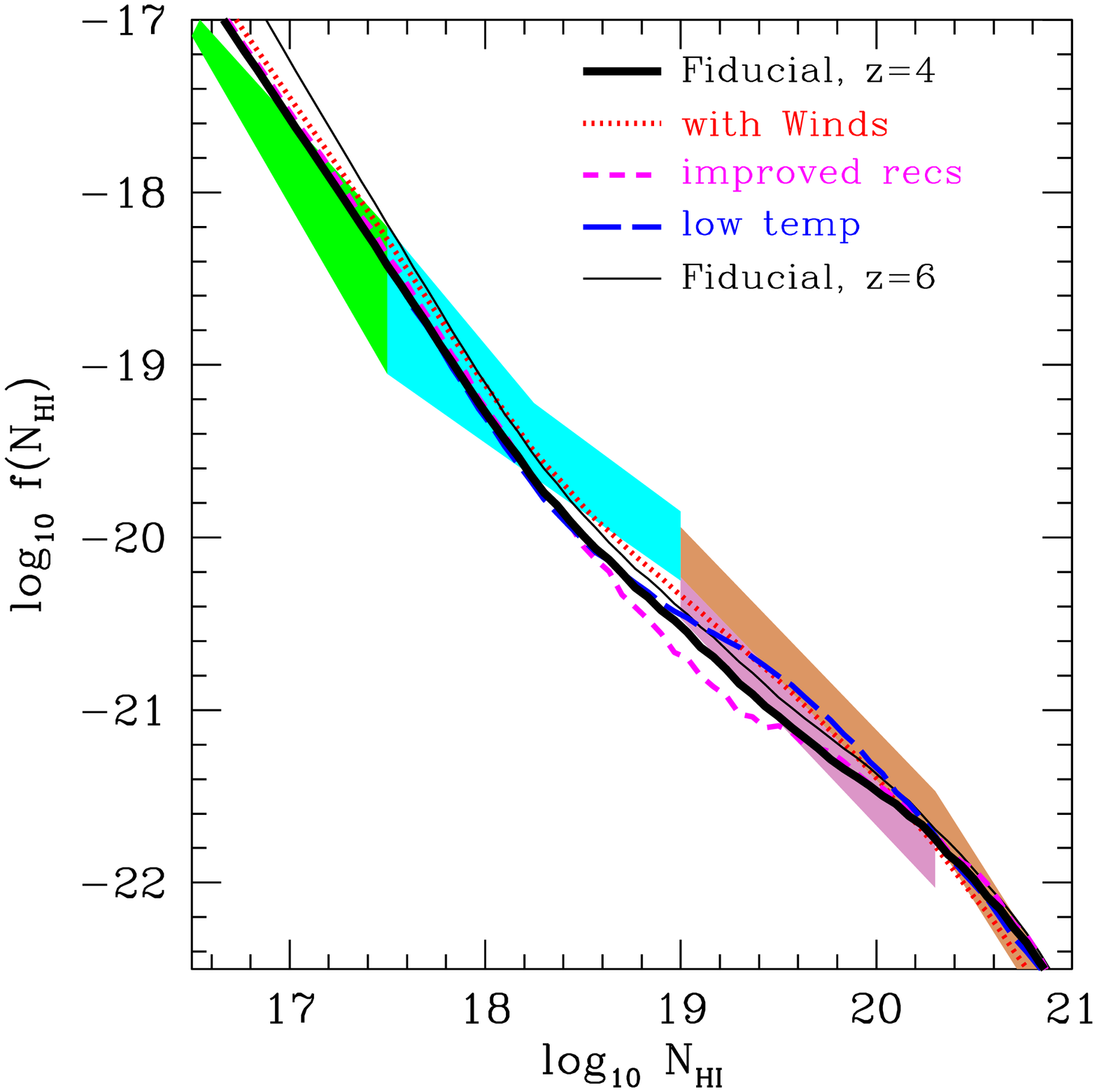, width=8.5cm}}
\end{center}
\caption{Comparison of $f(N_{\rm HI})$ in our calculations with observational constraints.  The shaded regions show the observational constraints at a mean redshift of $3.7$ \citep{prochaska09,omeara07,2009ApJ...696.1543P,prochaska10}.  Left panel:  The black thick solid curve is the fiducial calculation, which takes $\Gamma _{-12}=0.5$, the blue dashed curves are the same calculation but with $\Gamma_{-12}=0.25$ or $1$ (upper and lower curves, respectively), and the black thin solid curve assumes no attenuation of the ionizing background (i.e., optically thin radiative transfer).  Right panel:  The black thick solid curve is the fiducial calculation, the red dotted curve is from the simulation with galactic winds, the magenta short-dashed curve uses the alternative recombination prescription, and the blue long-dashed curve assumes a gas temperature of $10^4~$K in self-shielding regions.  The aforementioned curves in both panels are for $z=4$, and the black thin solid curve in the right panel is the same as the black thick solid but for $z=6$.   All curves in the right panel are calculated for $\Gamma _{-12}=0.5$.    The inset in the left panel shows the mean free path of $1~$Ry photons in proper Mpc.  The red error bars in the inset are the \citet{prochaska09} measurement (2$\sigma$ errors) and the magenta error bars are the \citet{songaila10} measurement assuming $\beta = 1.3$ (1$\sigma$ errors). \label{fig:compobs}\label{fig1}}
\end{figure*}

\subsection{Observations}
\label{ss:observations}

\citet{prochaska09} and \citet{prochaska10} recently presented measurements of $\partial^2{\cal N}/\partial z\,\partial N_{\rm HI}$ over the interval $10^{16} \lesssim N_{\rm HI}<10^{19}\,$cm$^{-2}$.  
The shaded regions in Figure \ref{fig:compobs} (excluding the pink region) show the constraints compiled in \citet{prochaska10} on
\begin{equation}
f(N_{\rm HI})\equiv\frac{\partial^2{\cal N}}{\partial z\,\partial N_{\rm HI}}\,\frac{H(z)}{H(0)\,(1+z)^{2}},
\end{equation}
where $H(z)$ is the Hubble parameter.
The function $f(N_{\rm HI})$ is independent of redshift if the number density and proper size of the absorbers does not evolve.  \citet{prochaska10} inferred that the power-law index of $f(N_{\rm HI})$, $-\beta$, must be steeper than $-1.7$ for $N_{\rm HI}<10^{17.5}~$cm$^{-2}$.  This bound constrains the density profile of overdensity $\sim 100$ systems, and we will show has interesting implications.  In addition, the red error bars in the inset in the left panel of Figure \ref{fig:compobs} show the novel constraint of \citet{prochaska09} on the mean free path of $1~$Ry photons, $\lambda$, obtained by stacking spectra. 

However, the column-density distribution of systems with $10^{16} \lesssim N_{\rm HI} < 10^{19}~$cm$^{-2}$ is not measured directly in \citet{prochaska10}.  Rather, it is inferred by interpolating between the constraints on (1) the number density of systems with $N_{\rm HI}>10^{17.5}~$cm$^{-2}$, (2) $\lambda$, (3) $f(N_{\rm HI})$ at columns probed by the Ly$\alpha$ forest, and (4) measurements at $N_{\rm HI} > 10^{19}~$cm$^{-2}$.  The specifics of this interpolation are described in \citet{prochaska10}.  

We note that \citet{omeara07} concluded that the normalization of the fit resulting in the tan region may be be biased high.  This study found a large excess of absorbers in the smallest $N_{\rm HI}$ bin, which could owe to spuriously identified systems.  When this bin was excluded from their analysis, it shifted the best-fit normalization of $f(N_{\rm HI})$ down by a factor of $\approx1.5$.  The pink region in Figure \ref{fig:compobs} assumes a normalization that is reduced by a slightly larger factor of $2$ (which turns out to be the factor needed to agree with our fiducial calculation; Section \ref{ss:sims}).  For this alternative normalization, the cyan region in Figure \ref{fig:compobs} should also be adjusted to enforce continuity.



\citet{songaila10} recently measured the abundance of Lyman-limit systems between $0 < z <6$, which extended this measurement to higher redshifts than reported in previous studies.  The magenta points with error bars in the inset in Figure \ref{fig:compobs} show this measurement, using $\beta= 1.3$ as was assumed in their study to extrapolate to smaller systems than were measured.  Using instead $\beta = 1.8$ would reduce the amplitude of these points by a factor of $0.3$.

%
 
\subsection{Simulations}
\label{ss:sims}

Almost all cosmological simulations take the intergalactic gas at $z\lesssim6$ to be in photoionization equilibrium with a homogeneous radiation field.  However, the approximation of a uniform ionizing background is not appropriate in systems that can self-shield to this radiation.  To allow for self-shielding, we post-process smooth particle hydrodynamics simulations that were run with the GADGET-2 code \citep{springel05} with a specialized radiative transfer algorithm. 

Our fiducial simulation uses a $512^3$ gas particles and the same number of dark matter particles in a box of size $20/h$~comoving Mpc (cMpc), resulting in dark matter and gas particle masses of $5.4\times10^6~\Msun$ and $1.1\times10^6~\Msun$ respectively.\footnote{Over and above our radiative transfer calculations, the ionizing background is turned off on-the-fly in these simulations for regions with number densities of hydrogen greater than $10^{-2}~$cm$^{-3}$ in order to mimic self-shielding as described in \citet{faucherlya}.  We find that this switch has little impact on our results, suppressing $f(N_{\rm HI})$ by a maximum of $20\%$ at $3\times 10^{18} < N_{\rm HI} < 3\times 10^{20}~$cm$^{-2}$ and having no effect at smaller columns.}  These simulations resolve the $\gtrsim 10^{9-10}~\Msun$ halos that are expected to retain their gas after reionization with thousands of particles and, thus, are the most likely hosts of self-shielded gas.  However, we have also checked for convergence with regard to cosmic variance and resolution by comparing with a $40/h$~cMpc simulation run with the same parameters and the same number of particles.\footnote{We find that the normalization of $f(N_{\rm HI})$ in the larger box is $10\%$ higher fairly uniformly for $N_{\rm HI} < 10^{20}~$cm$^{-2}$, while it undershoots by $10\%$ at $N_{\rm HI}\sim 10^{21}~$cm$^{-2}$.}   

In addition, we also use a $40/h~$cMpc, $2\times512^3$ simulation with a prescription for galactic winds to gauge the importance of such feedback on our results.  This simulation uses the wind implementation of \citet{springel03} with a mass loading factor of $2$ and with a wind velocity of $680~$km~s$^{-1}$.  In this wind implementation, twice the mass formed in stars is ejected in a biconical wind with approximately the said velocity, and the wind is briefly decoupled from the gas such that it can escape from the galaxy.   This simulation should be considered a relatively extreme model for winds based on the heavy suppression of the galaxy mass function it yields (see \citealt{faucherinprep}). 

Next, we use the SKID algorithm\footnote{\url{http://www-hpcc.astro.washington.edu/tools/skid.html}} to locate dense regions in the simulation by linking all particles within $0.5$ mean inter-particle spacings (with the additional criterion that the baryonic 
overdensity of associated particles is $>2.2$) and place each particle group onto a $120^3$ rectangular grid for the radiative post-processing.  The typical size size of this region is a few tenths of a cMpc.  We have run the same calculations on a $60^3$ grid and found excellent convergence.  We have also verified convergence with respect to the linking length choice.


Each group is illuminated with a power-law intensity per unit frequency with index $-\alpha$ from $1$ to $4~$Ry, and its amplitude is set by the \HI\ photoionization rate $\Gamma$.  In the fiducial calculation, we use $\alpha=1$ and $\Gamma_{-12}=0.5$, where $\Gamma_{-12}$ is $\Gamma$ in units of $10^{-12}\;$s$^{-1}$.  These choices are in line with measurements at $2<z<5$ and UV background models \citep{bolton07, haardt96, faucher08b, faucher09}.  We find that changing $\alpha$ by $\pm1$ alters $f(N_{\rm HI})$ with respect to the fiducial case by $<10\%$ at $z=4$, with the largest effect at columns of $10^{18} < N_{\rm HI} < 10^{20}~$cm$^{-2}$.  The radiative transfer algorithm casts six plane-parallel inward fronts of rays along the axes of the grid.  The rays are attenuated by the \HI\ photoelectric optical depth, and the intensity that reaches each cell sets the local \HI\ fraction under the assumption of ionization equilibrium and using the simulations' original gas temperatures.  The \HeI\ fraction is assumed to trace the \HI\ fraction with the rest of the helium in \HeII, which is a good approximation in \HI\ Lyman-limit systems \citep{mcquinnHeI}.  The algorithm then iterates to achieve convergence.  The assumptions of a power-law specific intensity and plane-parallel sources aligned with the grid allow us to use analytic expressions for the average photoionization rate in a cell as a function of the total \HI\ column to the cell and the \HI\ column across the cell itself.\footnote{The ionization front can span less than a proper kpc at relevant densities and, thus, is often not resolved in our calculations.  We find that iterating on the volume-averaged photoionization rate in a (uniform density) cell rather than, for example, the cell-centered value is critical to converge properly. }

\section{Comparison with Observations}
\label{sec:comparison}

\begin{figure}
\begin{center}
{\epsfig{file=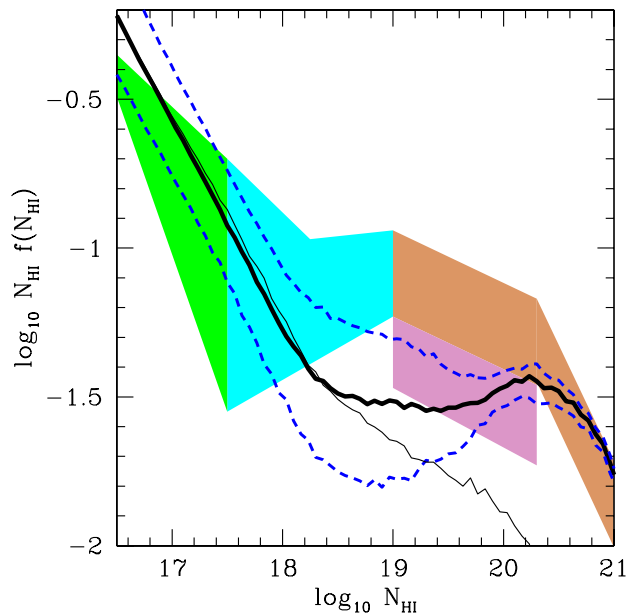, width=8.5cm}}
\end{center}
\caption{Same curves and shaded regions as in the left panel of Figure~\ref{fig1}, but the y-axis plots $N_{\rm HI} \, f(N_{\rm HI})$ rather than $f(N_{\rm HI})$. \label{figNFN}}
\end{figure}

The results of our fiducial calculation at $z=4$ are given by the black thick solid curve in both panels of Figure~\ref{fig:compobs}.  This curve has a similar slope at $N_{\rm HI}<10^{17.5}~$cm$^{-2}$ to what has been inferred observationally (green shaded region).  The flattening at higher columns, which physically occurs because of the onset of self-shielding \citep{katz96, zheng02}, is less prominent in the fiducial calculation than in the observations.  (The effect of this flattening in the fiducial calculation can be gleaned by comparing with the black thin solid curve, which assumes that the ionizing background is not attenuated.)  However, the fiducial model is consistent at $2\,\sigma$ with two numbers: the number density of systems with $N_{\rm HI}>10^{17.5}~$cm$^{-2}$ at $z=4$ and the mean free path of ionizing photons (see inset in the left panel of Fig. \ref{fig:compobs}), the two constraints that principally determine the green and cyan regions.  For a finer comparison, see Figure \ref{figNFN}, which instead plots $N_{\rm HI} \, f(N_{\rm HI})$ for the same quantities as in the left panel of Figure \ref{fig:compobs}.


The level of agreement with observations is an improvement over previous numerical studies.   \citet{katz96} and \citet{gardner01} found that Lyman-limit systems were underproduced by an order of magnitude in their cosmological simulations, likely because of the lower resolution of their calculations.  The simulations in \citet{gardner01} were run in a $11\,h^{-1}~$cMpc box with $2\times 64^3$ particles and had particle masses that were roughly two orders of magnitude larger than in our simulations.  Our results are in better agreement with those of \citet{kohler07}, who, for a suitable choice of $\Gamma$,  was able to reproduce the abundance of Lyman-limit systems in their $4\,h^{-1}~$cMpc and $8\,h^{-1}~$cMpc simulations with $128^3$ dark matter particles and the same number of Lagrangian gas elements.  However, our $f(N_{\rm HI})$ has more structure than the power law-like form found in \citet{kohler07}.  Our calculations qualitatively agree with those of Altay~et~al.\nocite{altay10} (2011, submitted concurrently with this study), which used simulations with similar box sizes and particle numbers to those discussed here.   


The form of $f(N_{\rm HI})$ in our calculations depends on several factors.  The thin solid curve in the right panel in Figure \ref{fig:compobs} is $f(N_{\rm HI})$ at $z=6$ for the fiducial background model.  This curve is similar to the $z=4$ curve, but is steeper at $N_{\rm HI}<10^{18}~$cm$^{-2}$.  In addition if the background is varied so that $\Gamma_{-12}=0.25$ or $1$ (which spans the range of most estimates; \citealt{faucher08b}), $f(N_{\rm HI})$ becomes respectively the upper and lower blue dashed curves in the left panel of Figure \ref{fig:compobs}.   The fiducial calculation does not incorporate winds that recycle gas from within galaxies.  The red dotted curve in the right panel is $f(N_{\rm HI})$ calculated from the same ionizing background model as the fiducial model curve, but with a simulation that incorporates galactic winds as described in Section \ref{ss:sims}.   The abundance of Lyman-limit systems is increased slightly in the case with winds compared to our fiducial calculation because winds result in more gas surrounding halos.  We have also looked at a simulation that assumes half of the wind velocity and half the mass loading factor as the aforementioned case and found surprisingly similar results compared to this model with more extreme winds. 

Our calculations assume Case A recombinations.  This neglects the reabsorption of ionizing photons produced by recombinations to the ground state.  The magenta short-dashed curve in the right panel in Figure \ref{fig:compobs} locally accounts for this reabsorption by using $\alpha_B+(\alpha_A-\alpha_B)\,\Gamma'/\Gamma$ as the recombination coefficient in a grid cell, where $\Gamma'$ is the local (attenuated) value of the photoionization rate, and $\alpha_A$ and $\alpha_B$ are the Case A and B recombination coefficients.  This interpolates between Case A and Case B recombinations in optically thin and thick regions, respectively.   This improved prescription suppresses the abundance of $N_{\rm HI} \sim 10^{19}$cm$^{-2}$ systems at the $10\%$ level.  A countering effect to neglecting reabsorption is that the simulations' temperatures are likely overestimated in self-shielding regions owing to suppressed atomic cooling (\S\ref{cooling}).  The blue long-dashed curve is for a toy temperature model where the temperature is set to $T = [10^4+(T_0-10^4)\,\Gamma'/\Gamma]~$K, where $T_0$ is the original temperature in the simulation.  Over relevant columns, $T_0  \approx 15-20\times 10^4~$K in the simulations so that $T \leq T_0$ in this model.  The primary difference between this model and the fiducial model is that $f(N_{\rm HI})$ is increased at $N_{\rm HI} \sim 10^{19.5}\,$cm$^{-2}$.  While this model likely overcompensates for the effect of this additional cooling, it does not account for the impact of lower temperatures on the gas dynamics (\S\ref{cooling}).  



\section{The case for a strong relationship between emissivity and $\Gamma$}
\label{sec:relationship}
We now focus on the implications of the functional form of $f(N_{\rm HI})$.  In particular, we show that its form implies that small changes in the emissivity of the sources can result in much larger changes in $\Gamma$.

\subsection{The \citet{miralda00} Model}
\label{ss:analytics}

We first describe the semi-analytic model of \citealt{miralda00} (henceforth MHR), which is useful for interpreting the observations and numerical calculations.  The MHR model assumes that there is a critical density $\Delta_i$ above which systems self-shield from the ionizing background and remain neutral.  We will show that our calculations support this assumption.  Given $\Delta_i$, the volume filling fraction of neutral gas, $F_V$, and the global recombination rate density, ${\cal R}$, can be calculated from just the volume-weighted gas density probability distribution $P(\Delta)$, a quantity easily measured from cosmological simulations.  For example, for isothermal gas ${\cal R} \propto \int_0^{\Delta_i} d\Delta \Delta^2 \, P(\Delta)$.  We will subsequently use that ${\cal R}$ was in balance with the ionizing emissivity of the sources, $\epsilon$, which is a good approximation at $z>2$ and after reionization (e.g., \citealt{madau99}).

This model further assumes that the mean free path at $1~$Ry is given by $\lambda=a_0\,F_V(\Delta_i)^{-2/3}$, where $a_0$ is an adjustable parameter.  This relation for $\lambda$ would hold if the comoving number density and shape of self-shielding regions were independent of $\Delta_i$.  The primary conclusion that we derive from the MHR model would be strengthened if the number density decreased or the shape became more spherical with $\Delta_i$, as one would anticipate.

Intuition is gained by approximating $P(\Delta)$ as a power law with index $-\gamma$ over the densities that contribute most of the recombinations \citep{furlanetto05}.  In this approximation, the MHR model yields the relations
\begin{eqnarray}
\lambda&=&~a_0\,\left[\int_{\Delta_i}^\infty d\Delta P(\Delta)\right]^{-2/3}\propto\Delta_i^{2 \,(\gamma-1)/3},\label{eqn:relationlambda}\\
\epsilon&=&~ {\cal R}~ = ~\alpha_{\rm A}\,\bar{n}_e\bar{n}_{\rm H} \;\int_0^{\Delta_i}d\Delta\,\Delta^2 P(\Delta)\propto\Delta_i^{3-\gamma},\label{eqn:relationA}\\
\Gamma &\approx&\frac{\sigma_{\rm LL}(3\beta-3+\alpha)}{3+\alpha}\epsilon\,\lambda  \propto\Delta_i^{(7-\gamma)/3}\propto\epsilon^{\frac{7-\gamma}{9-3\gamma}}\propto\epsilon^{1/(2-\beta)},\nonumber\\
\label{eqn:relationG}
\end{eqnarray}
where $n_X = \bar{n}_X\Delta$ is the number density in species $X$, $\sigma_{\rm LL}$ is the photoionization cross section at $1~$Ry, $-\beta$ is the power law index of $f(N_{\rm HI})$ at $N_{\rm HI}\approx10^{17}~$cm$^{-2}$, and we have assumed that the gas is isothermal (which holds approximately in our simulations at relevant densities).  Equation (\ref{eqn:relationG}) is valid for $\alpha>0$, and it assumes that cosmological expansion is unimportant (valid at $z\gtrsim2$) and that the mean free path of an ionizing photon with energy $E$ equals $\lambda\,[E/1~{\rm Ry}]^{3\beta-3}$ (e.g., \citealt{haardt96, miralda03}).  
The final proportionality in equation~(\ref{eqn:relationG}) assumes spherically symmetric power-law density profiles such that $\gamma$ can be related to the $\beta$ that applies for optically thin systems via the relations $f(N_{\rm HI})\,dN_{\rm HI}\propto 2\pi r \,dr$, $P(\Delta)\,d\Delta\propto 4\pi r^2 dr$, and $n_{\rm HI} \propto \Delta^2$ (i.e., there is an annulus at radius $r$ with column $N_{\rm HI}$ and a shell at $r$ with density $\Delta$).  In the previous section, we found that $\beta \approx 1.8$ for optically thin systems in both the simulations and in observations.  Equation (\ref{eqn:relationG}) indicates that such a $\beta$ results in a strong relationship between $\epsilon$ and $\Gamma$.  

\begin{figure}
\begin{center}
{\epsfig{file=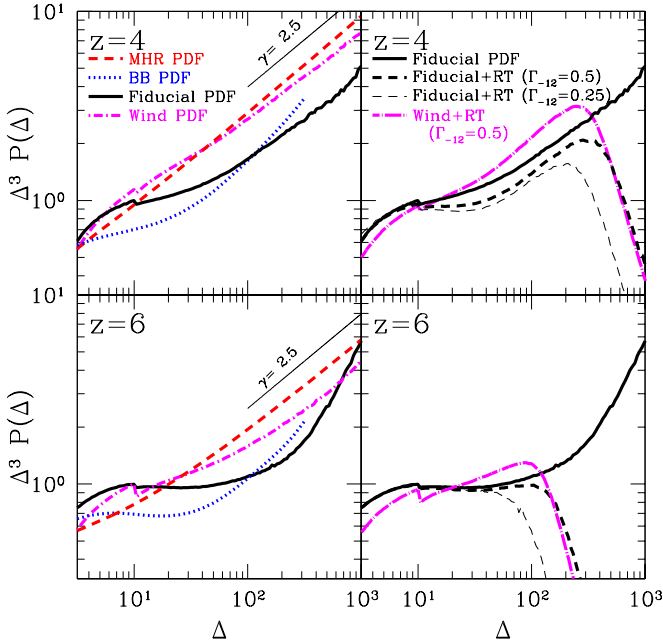, width=9cm}}
\end{center}
\caption{$\Delta^3$ times the PDF of the gas density, $P(\Delta)$, at the specified redshift.  The curves in the panels on the left are $\Delta^3\,P(\Delta)$ from published gas density PDFs and from our calculations.  The curves labeled ``RT'' in the panels on the right only include the contribution from photoionized gas as described in the text, which results in a cutoff at high densities.  The key in the upper panels applies to both the upper and lower panels. \label{fig:PDF} \label{fig2}}
\end{figure}

The power-law of the density PDF, $\gamma$, is more fundamental for determining this relationship than the power-law of $f(N_{\rm HI})$, $\beta$.  At high densities, the gas density PDF in the simulations can be approximated with the power-law $2.5 < \gamma < 3$, which also implies a strong scaling between $\epsilon$ and $\Gamma$ (eqn. \ref{eqn:relationG}).   The left panels in Figure \ref{fig:PDF} show $\Delta^3P(\Delta)$ at $z=4$ (top right panel) and $z=6$ (bottom right panel).\footnote{The small break at $\Delta = 10$ in the curves in Figure \ref{fig:PDF} originates because we are patching together $P(\Delta)$ from a calculation that uses a linking length of $1$ mean interparticle spacing with our fiducial calculation that uses a linking length of $0.5$ in order to cover the plotted range in $\Delta$.}  
The dashed red curve is the fit to $P(\Delta)$ in MHR.  The blue dotted curve is the fit to more recent simulations by \citealt{bolton09b} (BB).  
The black solid curve is $\Delta^3 P(\Delta)$ in the simulation without winds, and the magenta dot-dashed curve is the simulation with strong winds.  The simulation with winds has more gas around galaxies, which increases $\Delta^3P(\Delta)$ somewhat relative to the fiducial simulation.\footnote{The different $P(\Delta)$ curves should not be expected to agree for a few reasons.  These include that they were computed from simulations which had different thermal histories and because the MHR fit was to a simulation using an outdated cosmology and with a rigid parametrization.}  

Interestingly, all of the $P(\Delta)$ yield $\gamma\gtrsim2.5$, where $-\gamma$ is the power-law index of $P(\Delta)$.  For the MHR fit to $P(\Delta)$, $\gamma\approx2.5$ and equation (\ref{eqn:relationG}) implies that $\Gamma\propto\epsilon^n$ where $n=3$.\footnote{For power-law density profiles, $\gamma = 2.5$ corresponds to isothermal spheres with $\Delta \propto r^{-2}$, and $\gamma = 3$ corresponds to $\Delta \propto r^{-1.5}$.}  Thus, the intergalactic \HI\ photoionization rate scales as the \emph{cube} of the emissivity.  However, $n$ is extremely sensitive to $\gamma$ in this equation.  The $P(\Delta)$ in our simulations and in those of BB show $\gamma=2.5-3$ at relevant $\Delta$ (with $\gamma$ increasing with redshift), which formally imply $n=3-\infty$ given equation (\ref{eqn:relationG}).  Therefore, the flatness of $\Delta^3P(\Delta)$ indicates a strong relationship between $\epsilon$ and $\Gamma$. 

How justified is the MHR assumption of a critical density at which gas becomes neutral?
The curves labeled ``RT" in the right panels of Figure~\ref{fig:PDF} are the PDF of ionized gas in our calculations, or more specifically $\Delta^3P(\Delta)$ multiplied by  $\Gamma n_{\rm HI}/[\alpha_{\rm A}\,n_{H}^2]$.  The latter factor is equal to the square of the ionized fraction in the absence of collisional ionizations.  The recombination rate is proportional to the logarithmic integral over the RT curves.   The RT curves follow the density PDF curves before an abrupt cutoff, which supports the MHR model assumption of a characteristic $\Delta_i$.


\subsection{Numerical Calculations}
\label{numerics}

\begin{figure}
\begin{center}
{\epsfig{file=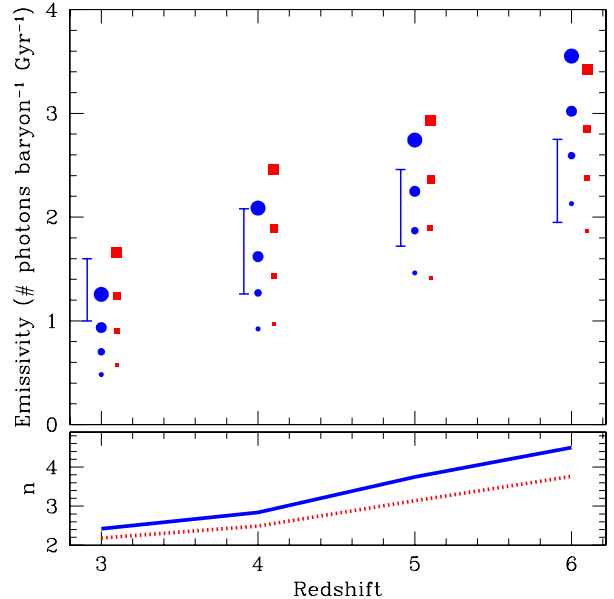, width=8.5cm}}
\end{center}
\caption{Top Panel: Number of \HI-ionizing photons emitted per hydrogen atom per Gyr as a function of redshift.  The blue circles are the fiducial calculation, and the red squares are the calculation with winds.  The different points correspond to $\Gamma_{-12}=0.1,~0.25,~0.5$, and $1$, with the point size increasing with $\Gamma_{-12}$.  The error bars bracket the observational constraints on $\Gamma_{-12}$, assuming the fiducial model mapping from $\Gamma_{-12}$ to $\epsilon$.  Bottom panel:  The power-law index that relates $\epsilon$ to $\Gamma$ from the fiducial (blue solid curve) and wind (red dotted curve) simulations.  These curves are computed from the ratio of the emissivity between the $\Gamma_{-12} =0.25$ and $0.5$ calculations.\label{fig:clumping} \label{fig3}}
\end{figure}

We can measure the power-law relationship between $\epsilon$ and $\Gamma$ directly in the simulations by calculating the recombination rate of the gas (which is in balance with $\epsilon$) as a function of $\Gamma$.\footnote{Or, we could equivalently calculate the rate of absorptions given $f(N_{\rm HI})$, and we have verified that this yields identical results.  
}  Recombinations that result from collisional ionizations are omitted by counting photo-ionizations and using that this is equal to the photoionization-induced recombination rate.  The individual points in Figure \ref{fig:clumping} are calculated directly from the fiducial simulation (blue circles) and the simulation with winds (red squares).  Each set of points at fixed $z$ corresponds to this calculation for $\Gamma_{-12}=0.1,~0.25,~0.5$, and $1$, with the point size increasing with $\Gamma_{-12}$.\footnote{Our group-finding method is $50\%$ complete at $\Delta = 5$ and more complete at higher $\Delta$.  We assume the MHR fit for $P(\Delta)$ and isothermal $20,000~$K gas to add the contribution to ${\cal R}$ from lower densities.  This correction is largest in the calculation at $z=6$ and $\Gamma_{-12} = 0.1$, where this gas contributes $25\%$ of the total recombinations.  Taking lower temperatures for this gas would result in more recombinations and would increase $n$ somewhat. 
}  

Our numerical calculations indicate that $\Gamma$ should scale strongly with emissivity.  At $z=3$, we find $n=2.2-2.4$ in the simulations (bottom panel, Fig. \ref{fig:clumping}).  At $z=4$, $n=2.5-2.8$, which is slightly weaker than the MHR model prediction for $\gamma=2.5$ of $n=3$.  Section \ref{ss:colion} provides an explanation for this weaker dependence.  At $z=6$, we find an even stronger scaling, with $n=3.8-4.5$.  Thus, a small change in $\epsilon$, particularly at $z=6$, could result in a much larger change in $\Gamma$.  For example, if $\epsilon$ changes from $2.1$ to $2.5~$photons per hydrogen atom per Gyr at $z=6$, then $\Gamma_{-12}$ evolves from $0.1$ to $0.25$.   We argue in Section \ref{sec:conclusions} that this behavior could explain the quick evolution in $\Gamma$ that is observed at $z\approx 6$ \citep{fan06}, which many studies have interpreted as indicating the overlap stage of reionization.  

These calculations also constrain the intergalactic emissivity history.
The error bars in Figure \ref{fig:clumping} bracket the blue points' $\Gamma_{-12}$ values that are consistent with measurements \citep{faucher08b, calverley10}.  In particular, they bracket $\Gamma_{-12} = \{0.6-1.4,~0.25-1,~0.2-0.7,~0.07-0.3\}$ at $z=\{3, 4, 5, 6\}$.   Each error bar generously spans a factor of $\approx4$ in $\Gamma_{-12}$, which is much larger than the error quoted in many studies and spans the range of most previous estimates.  However, since $\epsilon$ depends weakly on $\Gamma$ in these calculations, the uncertainty in the inferred emissivity is much smaller.

If $\Gamma_{-12}$ evolves from $\approx1$ at $z=3$ to $\approx0.1$ at $z=6$ as observations suggest, this evolution results in a slight increase in the comoving emissivity towards higher redshift in our calculations.  The primary difference between our weakly increasing comoving emissivity history and the weakly decreasing history of \citet{bolton07} [calculated using constraints on $\Gamma$ and the $\lambda$ calculated with the MHR model] is that our mean free path values decrease more rapidly with increasing redshift.  At $z=6$, a comoving emissivity of approximately two ionizing photons per hydrogen atom in the age of the Universe ($1~$Gyr) is needed in our calculations to be consistent with the constraints on $\Gamma_{-12}$.  This results in the same dilemma posed in \citet{bolton07} that this comoving emissivity was only sufficient to reionize the Universe by $z=6$ if $\epsilon$ were constant or increasing to higher redshift.  The increasing nature of our inferred comoving emissivity history with increasing redshift at $z>3$ may point to the solution.

These calculations also provide an estimate for the intergalactic clumping factor (the enhancement in the recombination rate over a homogeneous universe) since this number is equal to $\epsilon/[\alpha_A n_H^2]$.  Studies that estimate the ionizing emissivity from observations of $z>6$ galaxies need to know this uncertain factor to quantify whether this emissivity can keep the Universe ionized (e.g., \citealt{madau99, 2010Natur.468...49R}).  Unfortunately, there is a history of numerical simulations using very large subgrid clumping factors of $\sim 30$ to compensate for recombinations in ``unresolved'' structures.  (The number $30$ originated from counting gas that was ionized by local sources in the simulations of \citet{gnedin97} in the clumping factor tabulation.  The clumpiness of this gas, and hence the higher rate of recombinations in it, is better accounted for as an escape fraction of ionizing photons; \citealt{kohler07b}.)   Recent studies have calculated the clumping factor from simulations as a function of $\Delta_i$, finding $1-10$ (e.g., \citealt{pawlik09}).  Because our calculations determine the ionization structure of self-shielding regions, we can estimate the clumping factor directly rather than leave it as a function of $\Delta_i$.   In fact, the logarithmic integral over the RT curves in the right panel in Figure \ref{fig:PDF} equals the clumping factor.  We find a clumping factor enhancement of $2.4- 2.9$ at $z=6$ for the observationally allowed range of $\Gamma_{-12} = 0.1-0.25$ and using the simulation without winds (but see \S \ref{ss:rei}).  We find similar values using the simulation with winds.  Here we define the clumping factor as the enhancement in the recombination rate over a homogenous universe with a gas temperature of $10^4~$K.


\section{Additional Considerations}
\label{sec:other}

The results of this study rely on the density profile of $\Delta\sim\Delta_i$ systems being adequately simulated and the post-processing approximation for the radiative transfer.  
Here we address briefly how physical processes that are neglected in our calculations could alter the results.

 \subsection{Cooling Physics}
 \label{cooling}
The cooling of self-shielding gas is not properly captured in cosmological simulations without self-consistent ionizing radiative transfer (e.g., \citealt{faucherlya}).   In particular, the collisional cooling rate is proportional to the \HI\ fraction times the electron density, which is generally underestimated in the simulations at $N_{\rm HI}\gtrsim10^{17}~$cm$^{-2}$ since the temperature of the gas is computed assuming a homogeneous ionizing background.  (This error is more important than the cosmological simulations' \emph{underestimate} of the photoheating rate in self-shielding, photo-ionized gas.\footnote{It is only when systems are not highly photoionized -- $N_{\rm HI} \gtrsim 10^{19.5}~$cm$^{-2}$ in our calculations -- that the photoheating rate is overestimated in the simulations.}) Because collisional cooling is very efficient for gas with temperature greater than $\approx20,000~$K and inefficient at lower temperatures, underestimating the amount of collisional cooling may not lead to a serious error.  We find that if we set the temperature to $10^4~$K in self-shielding regions in post-processing as described in Section \ref{sec:comparison}, this operation flattens $f(N_{\rm HI})$ somewhat but does not affect $n$.  Lower temperatures and the correct ionization levels would also affect the distribution of the gas through their effect on the gas pressure.  See footnote~3 for a comparison that is relevant for estimating the significance of these effects.    


In addition, our simulations do not include metal line cooling.  \citet{prochter10} measured the metallicity of a Lyman-limit system and a few super Lyman-limit systems at $z\approx3.5$ to be $Z\approx10^{-1.7}~Z_\odot$, similar to the average metallicity of damped Ly$\alpha$ systems.  This metallicity is insufficient for metals to be the dominant coolant at $T\gtrsim20,000~$K, which requires $Z\gtrsim10^{-1}~Z_\odot$ \citep{wiersma09}.  However, this metallicity \emph{is} sufficient to allow ionized gas to cool to somewhat lower temperatures than in the simulations, where $\langle T\rangle\approx 20,000~$K at relevant columns.  
 
 

\subsection{Collisional Ionizations}
\label{ss:colion}
Feedback from galaxies could result in the gas within Lyman-limit systems being shock heated, driving it into a collisional rather than photoionization equilibrium.  If the Lyman-limit systems were in collisional ionization equilibrium, their cross section would be insensitive to $\Gamma$, so that $\Gamma$ would scale \emph{linearly} with the emissivity.  In our simulations with and without galactic winds, collisional ionizations are unimportant for $\Delta \sim 100$ gas at $z=6$, but become more important with decreasing redshift and, in fact, set the ionization state for half of $\Delta \sim 100$ gas at $z=3$.  In fact, we find that collisional ionizations play some role in making our calculations' $n$ smaller at $z\leq4$ than the predictions based on equation (\ref{eqn:relationG}).  


 As a final note, if $\beta$ in reality were as steep at $N_{\rm HI} \lesssim 10^{17.5}~$cm$^{-2}$ as the observations of \citet{prochaska10} suggest, collisional ionization is likely the only means to significantly weaken the dependence of $\Gamma$ on $\epsilon$ from the scaling in equation (\ref{eqn:relationG}).


\subsection{Proximity Effects}

While the extragalactic ionizing background almost certainly dominates in systems with $N_{\rm HI}< 10^{17.2}$cm$^{-2}$ \citep{miralda05}, local sources of radiation could be important for higher column-density systems \citep{schaye06}.  Our calculations do not account for their contribution.  The analytic model in \S \ref{ss:analytics} can be generalized to allow for a fraction of the gas at each density, $f(\Delta)$, to be ionized by local sources.  In this model, the $\Gamma$ dependence of the emissivity would only be weakened relative to our previous findings if $P(\Delta)\,[1-f(\Delta)]$ decreased less quickly with $\Delta$ than $P(\Delta)$.  However, the opposite seems more likely to hold, i.e., that local sources preferentially ionize the denser regions near the sites of star formation.  


\subsection{The Reionization History}
\label{ss:rei}
Reionization is only crudely included in our simulations, with the intergalactic gas being reionized homogeneously at $z\approx10$ in these calculations.  However, reionization could end as late as $z\approx 5.5$ \citep{mesinger10}.  Our results may be affected if reionization occurred at lower redshifts than is assumed in the simulations.  The relaxation timescale after a system was reionized is roughly equal to the Jeans length divided by the sound speed, or $H(z)^{-1}\,\Delta^{-1/2}$.  It takes $\Delta z\approx\Delta^{-1/2}\,(1+z)$ after a region was reionized for the gas to have relaxed.   For the $\Delta \sim 100$ systems of interest, this corresponds to $\Delta z\sim 0.1 \, (1+z)$.  After a large-scale region was reionized, the local $P(\Delta)$ would decrease, starting at the highest densities and evolving to lower densities.  This inside-out process would likely steepen the scaling between $\Gamma$ and $\epsilon$ a time interval of $\Delta z\gtrsim\Delta_i^{-1/2}\,(1+z)$ after a region was reionized.  These considerations would also result in our calculations underestimating the clumpiness of the ionized gas just after reionization, and thereby the emissivity, at fixed $\Gamma$.  
 
\section{Discussion}
\label{sec:conclusions}

The abundance of Lyman-limit systems in our fiducial calculation is consistent with observations at the $2 \,\sigma$ level.  We find that adding feedback from galactic winds or lowering $\Gamma$ improves the agreement.  This comparison tests whether our cosmological simulations reproduce the properties of gas at the outskirts of galactic halos.  However, our simulations may underproduce the number of $N_{\rm HI} \sim 10^{19}~$cm$^{-2}$ systems compared with observations.  If real, this discrepancy could indicate that the gas is cooler and, as a result, denser in these systems than in the simulations owing to additional cooling, or that the impact of feedback is more prominent than in the considered wind model.

Both observations and numerical simulations suggest a steep slope for $f(N_{\rm HI})$ with $\beta\approx1.8$ at $N_{\rm HI}\lesssim10^{17.5}~$cm$^{-2}$.  We showed that this steepness implies a strong relationship between the extragalactic \HI-ionizing background and the ionizing emissivity of the sources.  
At $z=3$, our calculations find that $\Gamma$ scales as $\epsilon$ to the power of $2.2-2.4$.  The near constancy of $\Gamma$ that is measured at $z\approx2-4$ \citep{bolton07, faucher08b} puts stringent requirements on the evolution of the total comoving ionizing emissivity owing to this strong scaling, requiring it to be roughly constant. This is an interesting finding since the ionizing background is believed to transition from being dominated by quasars to being dominated by star-forming galaxies over this redshift interval (e.g., \citealt{faucher08b}).


The scaling strengthens with redshift in our calculations, with $\Gamma$ proportional to $\epsilon$ to the $3.8-4.5$ power at $z=6$.  A strong scaling is related to the claim that the IGM clumping factor after reionization was small, independent of $\Delta_i$ \citep{pawlik09}, since both require $\gamma \approx 3$ for the high-density power-law index of the gas density distribution.  
Interestingly, \citet{fan06} detected a sharp decrease in the Ly$\alpha$ forest transmission at $z\approx6$.  If interpreted as a change in $\Gamma$, \citet{fan06} estimated that a factor of at least $2$ change was required over $\Delta z\approx0.5$.   It does not seem plausible that the emissivity of galaxies changed by such a large factor over this interval, which constitutes $7\%$ of the Hubble time at $z=6$.  Our simulations find that a more modest $\approx 20$\% change in the comoving emissivity at $z\approx6$ over $\Delta z\approx0.5$ would result in a factor of $2$ change in $\Gamma$.  Estimates for the variance in $\Gamma(z)$ between widely separated regions suggest that this effect could operate on this timescale \citep{mesinger09}.   

The quick evolution in $\Gamma$ from the mechanism discussed here is not necessarily associated with the end of reionization, defined as the epoch when neutral patches were present in the low-density IGM so that $\Delta_i \sim 1$.   In all of our calculations, $\Delta_i \gg 1$, with $\Delta_i \sim 50$ at $z\approx 6$ and for $\Gamma = 10^{-13}~$s$^{-1}$ -- approximately what is measured at this $z$ using the Ly$\alpha$ and Ly$\beta$ forests.  In addition, the mean free path of $1~$Ry photons is $8~$cMpc for this case.  This is still larger than the $0.5~(1.1)~$cMpc average separations of galaxies residing in $>10^8$ ($>10^9$)~$M_\odot$ halos in the assumed cosmology.  Therefore, it does not satisfy the simplest criterion for reionization that the mean free path was shorter than the separation of the anticipated sources (e.g., \citealt{miralda00}).\footnote{The Ly$\gamma$ forest measurements of \citet{fan06} suggest an even larger drop off in $\Gamma$ to $10^{-14}~$s$^{-1}$.  At this $\Gamma$, the mean free path in our calculations is $\sim 1~$cMpc, comparable to the distance between sources and difficult to reconcile with being post-reionization.  However, the $z\approx 6$ Ly$\gamma$ measurements derive from $\sim 30~$cMpc snippets from three sightlines, and their interpretation is sensitive to contamination from other Lyman-series transitions, uncertainties in the low density gas PDF, and the temperature-density relation \citep{furlanetto09}.}
 
The leading theory for explaining the rapid evolution in $\Gamma$ had been that it owed to the overlap of large-scale cosmological \HII\ regions (e.g., \citealt{gnedin00}).  However, it appears unlikely that overlap can produce such rapid evolution in $\Gamma$ for two reasons.  First, models predict large temporal variance between when different $\sim 30~$cMpc locations in the IGM were ionized, with the majority of the gas ionized over an interval of $\Delta z \gtrsim 3$ \citep{furlanetto04, mcquinn07}.  Second, by the time of overlap, it is likely that $\Gamma$ was regulated by absorptions in Lyman-limit systems rather than in diffuse \HI\ gas \citep{furlanetto09b}.  If this were the case, the overlap process itself could not mediate a sharp jump in $\lambda$ (and hence $\Gamma$), and fast temporal evolution in $\Gamma$ could only owe to the type of effect discussed here.\footnote{Alternative (lognormal) models for $P(\Delta)$ have also been invoked to explain this evolution \citep{becker07}.  We do not consider such explanations because, in the context of the concordance cosmology, one does not have this freedom.}

Our fiducial calculations favor the comoving emissivity of \HI-ionizing photons to have increased slightly with redshift from $z=3$ to $6$.  This trend is consistent with the conclusion that the emissivity had to be no more than weakly decreasing with redshift at $z>3$ to be able to reionize the Universe \citep{miralda03, bolton07, faucher08b}.  The improvement in this work over these previous studies is a self-consistent model for $\lambda(\Gamma)$, which is necessary to infer the emissivity at $z>4$.  We also showed that this conclusion is robust to the considerable uncertainty in $\Gamma$ at $z\approx 6$.  A growing emissivity with redshift could arise from the efficiency of escape of ionizing photons increasing with redshift or as a result of thermal feedback on low-mass galaxies.

In contrast with this result, simulations of reionization (e.g., \citealt{iliev06, trac07, zahn07}) generally prescribe an emissivity history that is strongly decreasing with increasing redshift \citep{choudhury09}.  
In addition, previous radiative transfer simulations of reionization in large enough box sizes to study the statistics of this process ($\gg10~$comoving Mpc; e.g., \citealt{iliev06, trac07, zahn07}) did not resolve the Lyman-limit systems and, thus, could not capture the effects discussed in this paper.  To capture these systems at $z\gtrsim 6$ necessitates resolving the ionization structure of gas parcels with $\Delta \sim 100$.  While this would require a large increase in resolution beyond that achieved in these previous simulations of reionization, a first step would be to incorporate Lyman-limit systems with a subgrid model based on the calculations described here. 
\\  
\\

We would like to thank Gabriel Altay, Lars Hernquist,  Dusan Keres, and Joop Schaye for useful discussions and/or for their help with the simulations.  We thank Adam Lidz, Xavier J. Prochaska, and especially Jordi Miralda-Escud{\'e} for comments on an earlier draft.  MM and SPO thank the Aspen Center for Physics for their hospitality.  MM is supported by the NASA Einstein Fellowship.  SPO acknowledges NSF grant AST~0908480.  CAFG is supported by a fellowship from the Miller Institute for Basic Research in Science and NASA grant 10-ATP10-0187.

\bibliographystyle{apj}
\bibliography{lymanlimits}

\appendix

\section{A. Properties of Lyman-limit Systems}
\label{ap:properties}

\begin{figure}
\begin{center}
{\epsfig{file=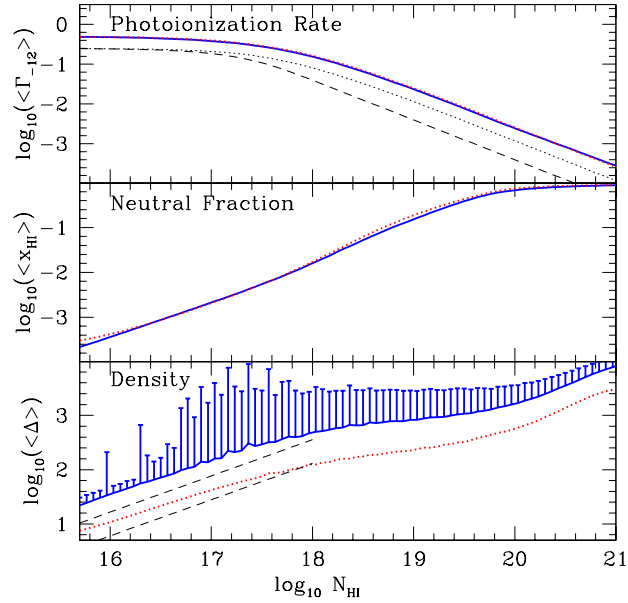, width=8.5cm}}
\end{center}
\caption{Average properties of absorbers as a function of $N_{\rm HI}$ in the fiducial calculations.  The solid blue curves are for $z=4$, and the red dotted curves are for $z=6$.  Each curve is calculated by weighting by $n_{\rm HI}$ along a skewer and then averaging skewers at fixed $N_{\rm HI}$.  The top panel shows the average of $\Gamma_{-12}$ within absorbers.  For comparison, the two black thin curves in the top panel are the same quantities but for the two analytic slab models described in Appendix \ref{ap:properties}.   The middle panel shows the average \HI\ fraction.  The bottom panel shows the average density in units of the cosmic mean, and the one-sided error bars shown for the $z=4$ case are the standard deviation in $\Delta$.  For comparison, the black thin dashed lines in the bottom panel represent the analytic model of \citet{schaye01} at $z=6$ and $z=4$ (lower and upper lines, respectively).      \label{fig:prop}}
\end{figure}

The panels in Figure \ref{fig:prop} show the average of the \HI\ photoionization rate, the fraction of hydrogen in \HI, and the density.   Each property is tabulated by weighing by $n_{\rm HI}$ along a skewer through a radiatively post-processed group and then averaging over all skewers through all groups at fixed $N_{\rm HI}$.  In each panel, the thick blue curve is the fiducial model at $z=4$, and the red dotted curve is the same but at $z=6$.  

The top panel shows the average of $\Gamma_{-12}$ as a function of $N_{\rm HI}$.   For comparison, the black thin dashed and dotted curves in this panel are respectively the average of $\Gamma_{-12}$ in a slab with column $N_{\rm HI}$ exposed to a planar source radiating orthogonal to the slab's plane and with (1) a monochromatic ionizing background at $1~$Ry (thin dashed curve) and (2) the same ionizing background as in the fiducial model (thin dotted curve).  Both curves assume $\Gamma_{-12} = 0.25$ to be incident on the slab in order to separate these curves from the simulation curves which took $\Gamma_{-12} = 0.5$.  The attenuation of $\Gamma$ in slab model (2) is almost identical to that in the simulations.  Both decrease as $N_{\rm HI}^{-1}$ at large columns.

The middle panel shows the average \HI\ fraction as a function of $N_{\rm HI}$.   Systems with $N_{\rm HI} \lesssim 10^{19.5}~$cm$^{-2}$ are highly ionized in our calculations.  This is consistent with observations, which find that Lyman-limit systems are highly ionized \citep{prochaska99} and the results of \citet{kohler07}.  It occurs because (1) harder photons require larger $N_{\rm HI}$ to be attenuated (a $4~$Ry photon has optical depth unity for $N_{\rm HI} = 1\times 10^{19}~$cm$^{-2}$)  and (2) even $\Gamma_{-12} \ll 1$ can be sufficient to keep the gas highly ionized \citep{zheng02}.  In our calculations and those of \citet{kohler07}, systems with $10^{17} < N_{\rm HI} < 10^{19}~$cm$^{-2}$ are significantly more ionized than those in the study of \citet{altay10}.  For example, systems with $N_{\rm HI}  =10^{18}~$cm$^{-2}$ in \citet{altay10} had a median \HI\ fraction of $0.1-0.3$  versus $10^{-2}$ here.

The bottom panel in Figure \ref{fig:prop} shows the average of $\Delta$.  The density of absorbers increases as $\approx N_{\rm HI}^{0.6}$ before plateauing over $10^{18} \lesssim N_{\rm HI} < 10^{20}$cm$^{-2}$ owing to self-shielding.  This plateau occurs at $\Delta \sim 500$ for $z=4$ and at $\Delta \sim 100$ for $z=6$.  The one-sided error bars represent the standard deviation in the density at $z=4$.  Prior to self-shielding, there is a tight relationship between density and $N_{\rm HI}$ because the absorbers' size is roughly the Jeans length and their ionization state is set by the amplitude of the ionizing background \citep{schaye01}.  Once self-shielding becomes significant, this relationship is broken.  As a result, the scatter becomes much larger at $N_{\rm HI} \gtrsim 10^{17}$cm$^{-2}$.  This scatter decreases again when the systems become neutral at $N_{\rm HI} \gtrsim 10^{20}~$cm$^{-2}$ as a similar relationship is again established.  Lastly, if we had plotted the average proper density rather than the density in units of the mean in the bottom panel, the $z=4$ and $z=6$ curves would almost overlap.

The black thin dashed lines in the bottom panel of Figure \ref{fig:prop} are the predicted relationship in the model of \citet{schaye01} between $\Delta$ and $N_{\rm HI}$ for optically thin systems with $\Gamma_{-12} = 0.5$ and $2\times10^4~$K gas (although, the temperature dependence in this model is weak).  This model just assumes photoionization equilibrium with the average $\Gamma$ and that the absorbers width is the Jeans length,  predicting the scaling $\Delta \propto N_{\rm HI}^{2/3}$.  We find that this simple model underpredicts $\Delta$ by a factor of $\approx 2$ at $N_{\rm HI} \approx 10^{16}~$cm$^{-2}$, with this difference decreasing with increasing $N_{\rm HI}$.  However, this model should not be expected to capture the normalization at better than the factor of $2$ level, and, more importantly, we find that it is successful at capturing the trend with $N_{\rm HI}$.  


\end{document}